\documentclass[german]{article}

\usepackage{amsmath}
\usepackage{amsthm}
\usepackage{amssymb}
\usepackage{amscd}
\usepackage{epsf}
\usepackage{rotating}
\usepackage{wrapfig}

\chardef\bslash=`\\ 

\makeatletter
\def\verbatim{\interlinepenalty\@M \@verbatim
  \leftskip\@totalleftmargin\advance\leftskip2pc
  \frenchspacing\@vobeyspaces \@xverbatim}
\makeatother
\theoremstyle{plain}

\theoremstyle{remark}

\numberwithin{equation}{section}

\def\1I{\relax{\rm 1\kern-.25em \rm l}} 

\def\Rahmen#1#2#3 {
   \vbox{\hrule height#2
         \hbox{
               \vrule width#2
               \hskip#1
               \vbox{
                     \vskip#1{}
                     \hbox{#3}
                     \vskip#1
                    }%
               \hskip#1
               \vrule width#2
              }
         \hrule height#2
        }}
\def\href#1#2{#2}

\begin{document}
\thispagestyle{empty}
\rightline{QMW-PH-00/14}
\rightline{hep-th/0011160}
\vspace{2truecm}
\centerline{\bf \Large A power law for the lowest eigenvalue in}
\vspace{0.4truecm}
\centerline{\bf \Large  localized massive gravity}
\vspace{2.5truecm}

\centerline{\bf \hspace{1ex}Andr\'e Miemiec\footnote{a.miemiec@qmw.ac.uk}}
\vspace{-.2truecm}
\begin{center}
{\em
      Department of Physics\\ 
    Queen Mary and Westfield College\\
            Mile End Rd\\ 
       London E1 4NS, UK
}
\end{center}

\vspace{2.0truecm}
\begin{abstract}
\noindent
This short note contains a detailed analysis to find the 
right power law the lowest eigenvalue of a localized massive 
graviton bound state in a four dimensional AdS background 
has to satisfy. In contrast to a linear dependence of the 
cosmological constant we find a quadratic one.  
\end{abstract}
\bigskip \bigskip
\newpage

\section{Definition of the Problem}
\label{SectionIntro}

The consistent embedding of four dimensional physics, in 
particular four dimensional gravity, into a higher dimensional 
spacetime progresses significantly, when the authors of   
\cite{Randall:1999vf} observed that the high dimensional 
graviton can be localized within the extra dimensions.
This opens a door for realizing the observable physics on 
a hypersurface in higher dimensional spacetime.  
After this discovery several attempts were made to generalize this 
construction to four dimensional spacetime with a non vanishing 
cosmological constant. The relevant solutions are constructed in 
\cite{Kaloper:1999sm,Kim:1999ja,Nihei:1999mt,DeWolfe:1999cp} and 
include, beside the original Minkowski space, de Sitter (dS$_4$) 
and anti de Sitter (AdS$_4$) spacetimes.
In the papers \cite{Kogan:2000uy} and \cite{Karch} the authors focus 
independently on the localization of massive gravity fluctuations 
around the cosmological AdS brane solution, i.e. they investigate the 
perturbation $h_{ij}$ around an AdS$_4$ background $g_{ij}$. The metric 
reads\footnote
{
  My conventions follow \cite{Karch}. There the action is given 
  this metric belongs to.
}:
\begin{eqnarray*}
  ds^2 ~=~ e^{2A(x)}\,\left[\vbox{\vspace{2.5ex}}\right.\,   
                             \left(\vbox{\vspace{2ex}}\right.\,
                                      g_{ij}\,+\,h_{ij}\,
                             \left.\vbox{\vspace{2ex}}\right)\,
                             d\xi^id\xi^j~-~ dx^2\,
                      \left.\vbox{\vspace{2.5ex}}\right]
\end{eqnarray*}
and the warp factor $A(x)$ in the conformal coordinate $x$ is given by
\begin{eqnarray}\label{WarpFactor} 
    A(x) ~=~ \ln\,\frac{ 
                            L\sqrt{-\Lambda}
                       }
                       {
                            \sin\,\left[\,
                                           \sqrt{-\Lambda}
                                           \left(\,|x|\,+\,x_0\,\right)\,
                                  \right]
                       }.
\end{eqnarray}
The contribution of the fifth dimension to the lowest order fluctuations 
of the gravitational field \cite{Brandhuber:1999hb,Csaki:2000fc} is encoded 
in the next 
equation
\begin{eqnarray}\label{Dgl}
    \left[\,
           -\,\frac{d^2}{d\,x^2} \,+\, \frac{9}{4} A'(x)^2 
                                 \,+\, \frac{3}{2} A''(x)\,
    \right]\,\psi(x) ~=~ \underbrace{\;m^2}_{E}\,\psi(x)
\end{eqnarray} 
and we are left with the problem to determine the spectrum of this 
differential operator and in particular the quantitative behaviour 
of the lowest eigenvalue $E_0$. This is precisely the question we are 
concerned with in this note. See \cite{Student} for numerical 
investigations of the same question and the similar results of 
\cite{Kogan:2000vb} in a slightly modified setting.\\ 
It is well known that such an operator can be factorized by aiding 
what is called the superpotential $W$. It is simply given through 
the warp factor $A(x)$ and reads  $W\,=\,\frac{3}{2}\,A'(x)$.
Introducing the operators $A$ and $A^\dagger$ below\footnote
{
  Since the warp factor does not appear anymore it is unambiguous to 
  use the symbol $A$ for the operators.   
}
\begin{eqnarray*}
   A \,=\,\phantom{-}\frac{d}{dx}+W,\hspace{7ex}
   A^\dagger \,=\, -\frac{d}{dx}+W
\end{eqnarray*}
the differential equation (\ref{Dgl}) can be written as 
$A A^\dagger\,\psi\,=\,m^2\psi$. 
We want to remind that this type of problem associate 
the spectra of two Hamiltonians \hbox{$H_1\,=\,A^\dagger A$} and 
\hbox{$H_2\,=\,A A^\dagger$} to each other as shown in 
Fig.~\ref{figureSpektren}. Both Hamiltonians are of the form 
\hbox{$H\,u\,=\,u''\,+\,V(x)\,u$} and the potentials generated by $W$ 
read:
\begin{eqnarray*}
        W ~=~ -\,\frac{3}{2}\,\alpha\,{\rm sgn}\,x\,\cot(\alpha |x|+\beta) 
               \hspace{2ex}
              \begin{array}{cl}
                  \nearrow   &   V_1 ~=~ \;~\;~\;
                                         3\,\alpha\,\cot\beta\;\delta (x)
                                         \,+\,\alpha^2
                                         \left(\,\frac{3}{4}
                                            \frac{1}{\sin^2(\alpha |x|\,+\,
                                                      \beta)} ~-~ \frac{9}{4}
                                         \,\right)\\ 
                             &                       \\
                  \searrow   &   V_2 ~=~ -\,3\,\alpha\,\cot\beta\;\delta (x)
                                         \,+\,\alpha^2
                                         \left(\,\frac{15}{4}
                                            \frac{1}{\sin^2(\alpha |x|\,+\,
                                                      \beta)} ~-~ \frac{9}{4}
                                         \,\right)             
              \end{array}\\
\end{eqnarray*}
Here $\,\alpha\,=\,\sqrt{-\Lambda}\,$ and 
$\,\beta\,=\,\sqrt{-\Lambda}\, x_0\,$ 
are convenient abbreviations used in the analysis.
The spectrum we are primarily concerned with belongs to the Hamiltonian 
$H_2$.\\
\parbox{\textwidth}{\vspace{0.2cm}
  \refstepcounter{figure}
  \label{figureSpektren}
  \begin{center}
  \begin{turn}{0}
  \makebox[5cm]{
     \epsfxsize=5cm
     \epsfysize=5cm
     \epsfbox{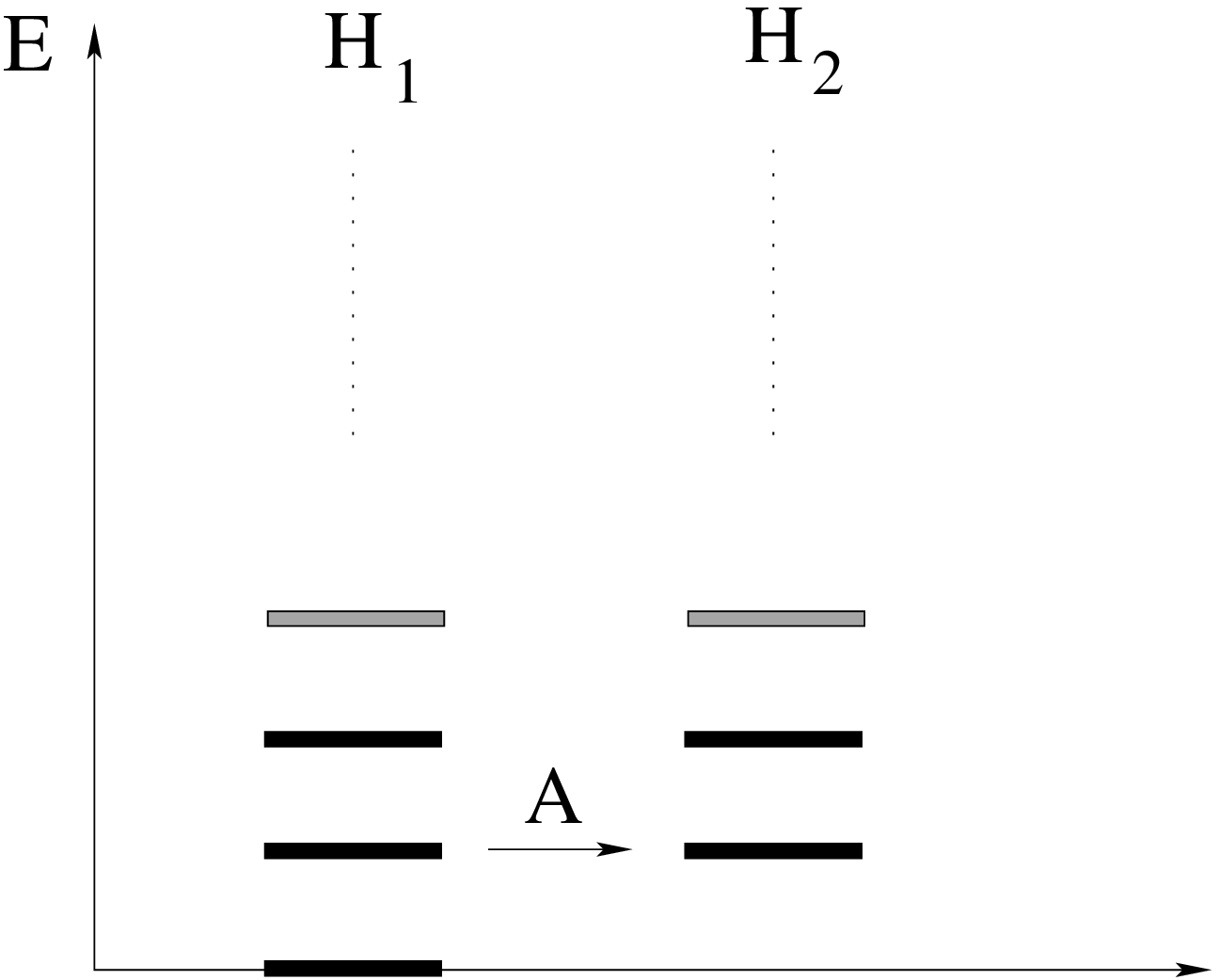}
  }
  \end{turn}
  \end{center}
  \center{{\bf Fig.{\thefigure}.} Spectra of $H_1$ and $H_2$}
}\\[1ex]
The additional structure present by this factorization is very useful 
for constructing the solutions. On the one side for special parameters,
$W$ depend on, it becomes possible to solve the Hamiltonian purely 
algebraically, i.e. the potential becomes self similar \cite{Cooper:1995eh}. 
For the potentials at hand the mismatch is parametrized by $\beta$. Only 
if $\beta\,=\,\pi/2$ they display this property in full beauty.\\        
On the other side a nice property of supersymmetric quantum mechanics is 
the fact that for parity conserving potentials the operator $A$ intertwines 
the parity and maps even to odd eigenfunctions of the dual problem 
and vice versa. Since we are interested in the lowest eigenvalue of the 
operator $H_2$, the eigenfunction of which is an even one, we can translate 
it into the problem to find the second eigenvalue of $H_1$, the eigenfunction 
of which is an odd one. 
The advantage of this is that we can neglect the whole trouble 
arising from the $\delta$-function. Since an odd eigenfunction is zero 
at the origin the product $\delta(x)\,\psi(x)\,=\,\delta(x)\,\psi(0)$ does 
not contribute and drops out. For this reason we can restrict our 
considerations to the potential below without the $\delta$-term.    
\begin{eqnarray*}
 \tilde{V}_1 ~=~ \alpha^2\;\left(\,
                                   \frac{3}{4}
                                   \frac{1}{\sin^2(\alpha |x|\,+\,\beta)} 
                                   ~-~ \frac{9}{4}
                           \,\right)\\ 
\end{eqnarray*}
This potential stays finite only in the interval 
\begin{eqnarray*}
     -\pi+\beta\,\leq~\alpha\,x~\leq\,\pi-\beta
\end{eqnarray*}
which forces us to require a vanishing wave function at the boundaries.
A typical picture the potential look like is given in 
Fig.~\ref{figurePotBeta05}.
Since we want to consider the odd solutions of the problem above, we also 
require the vanishing of the wave function at the origin. This leads by 
symmetry to the boundary conditions below: 
\begin{eqnarray}\label{Randbedingung}
 \psi^{(1)}(0) ~=~ \psi^{(1)}(\frac{\pi}{\alpha}-\frac{\beta}{\alpha}) ~=~ 0.
\end{eqnarray}
The full solution of the problem will be obtained stepwise. 
The first one is the solution of the differential equation for 
$\beta\,=\,0$. The second is to justify the arguments in a way, that the 
shifted potential coincides on the positive $x$-axis with the given 
potentials of interest. Furthermore we implement the  boundary conditions
for the odd solutions of $H_1$. In the last step we transform the odd 
solution of $H_1$ to an even solution of $H_2$.\\
After all we discuss the validity of an expansion of the lowest eigenvalue 
$E_0$ of $H_2$ in terms of $\beta$.

\section{The Power Law}

\subsection{The Differential Equation}
\label{SubSectionDGL}

After some mild manipulations including a rescaling of $x$ corresponding 
to $\xi = \alpha x$ and renaming $\xi$ to $x$ afterwards we arrive at the 
problem to find the solutions of: 
\begin{eqnarray*}
     -\,\frac{d^2u(x)}{dx^2} ~+~ 
        \left(
                \frac{3}{4}\frac{1}{\sin^2x}-\frac{9}{4}
        \right)\cdot u(x)
        ~=~ E\,u(x)
\end{eqnarray*}
Here $E$ stands for $E/\alpha^2$ as an side effect of this rescaling.\\  
    
\noindent
Introducing the new variable $y=\cos^2\,x$ and rescaling the resulting 
function  like $u(y)\,=\,\frac{v(y)}{(1-y)^{1/4}}$ we obtain
\begin{eqnarray*} 
   4\,y\,(y-1)\, v'' ~+~ 2\,(y-1)\,v' ~-~ (E+2)\,v(y) ~=~ 0.
\end{eqnarray*}
This is the hypergeometric differential equation. Since the argument 
$y$ varies along $0\leq y\leq 1$ the solution can be written as: 
\begin{eqnarray*}
    v(y) &=& c_1\,F(\,[\,
                         -\,\frac{1}{4}~+~\frac{\sqrt{4E\,+\,9}}{4},\,
                         -\,\frac{1}{4}~-~\frac{\sqrt{4E\,+\,9}}{4},\,
                      ],\,
                      [\,
                          \frac{1}{2}\,
                      ],\, y\,
                   )\\
         &+& c_2\,\sqrt{y}\,
                  F(\,[\,
                         \frac{1}{4}~+~\frac{\sqrt{4E\,+\,9}}{4},\,
                         \frac{1}{4}~-~\frac{\sqrt{4E\,+\,9}}{4},\,
                      ],\,
                      [\,
                          \frac{3}{2}\,
                      ],\, y\,
                   )\\
\end{eqnarray*}
Transforming back to the original problem we now obtain:
\begin{eqnarray*}
    u(x) &=& c_1\,\frac{1}{\sqrt{\sin\,x}}\,
                  F(\,[\,
                         -\,\frac{1}{4}~+~\frac{\sqrt{4E\,+\,9}}{4},\,
                         -\,\frac{1}{4}~-~\frac{\sqrt{4E\,+\,9}}{4},\,
                      ],\,
                      [\,
                          \frac{1}{2}\,
                      ],\, \cos^2\,x\,
                   )\\
         &+& c_2\,\frac{\cos\,x}{\sqrt{\sin\,x}}\,
                  F(\,[\,
                         \phantom{-\,}
                         \frac{1}{4}~+~\frac{\sqrt{4E\,+\,9}}{4},\,
                         \phantom{-\,}
                         \frac{1}{4}~-~\frac{\sqrt{4E\,+\,9}}{4},\,
                      ],\,
                      [\,
                          \frac{3}{2}\,
                      ],\, \cos^2\,x\,
                   )\\
\end{eqnarray*}

\subsection{The odd solutions of $H_1$}

After shifting the argument like $x\longrightarrow x\,+\,\beta$ we have   
to implement the boundary conditions of eq.~(\ref{Randbedingung}). The 
first of the two in eq.~(\ref{Randbedingung}) leads to the following choice 
for the constants $c_1$ and $c_2$: 
\begin{eqnarray*}
  c_1 &=& \phantom{-}\frac{\cos\,\beta}{\sqrt{\sin\,\beta}}\;
                       F(\,[\,\phantom{-\,}
                              \frac{1}{4}~-~\frac{\sqrt{9+4E}}{4},\, 
                              \phantom{-}
                              \frac{1}{4}~+~\frac{\sqrt{9+4E}}{4}\,
                           ],\,
                           [\,
                                \frac{3}{2}\,
                           ],\,\cos^2\,\beta\,
                        )\\
  c_2 &=& -\,\frac{1}{\sqrt{\sin\,\beta}}\;
                       F(\,[\,
                             -\frac{1}{4}~+~\frac{\sqrt{9+4E}}{4},\,
                             -\frac{1}{4}~-~\frac{\sqrt{9+4E}}{4}\,
                           ],\,
                           [\,
                               \frac{1}{2}\,
                           ]\,,\cos^2\,\beta\,
                        )
\end{eqnarray*}
The second boundary condition finally produces the transcendental 
equation, which determines the eigenvalues. To do that we have to 
keep in mind that the hypergeometric function evaluated at the point 
1 can be expressed through the  $\Gamma$-function as shown below:
\begin{eqnarray*}
    F(\,[\,a,b\,],[\,c\,],\,1\,)~=~
    \frac{\Gamma(c)\,\Gamma(c-a-b)}{\Gamma(c-a)\,\Gamma(c-b)}
    \hspace{1cm}\hbox{if}\hspace{0.5cm} c-a-b\,>\,0
\end{eqnarray*}   
Then one obtains:
{\small
\begin{eqnarray}
       0 &=&    \frac{\cos\beta}{
                                  \Gamma(\frac{3}{4}-\frac{\sqrt{4E+9}}{4})
                                  \Gamma(\frac{3}{4}+\frac{\sqrt{4E+9}}{4})
                                }
                       F(\,[
                              \frac{1}{4}-\frac{\sqrt{9+4E}}{4}, 
                              \frac{1}{4}+\frac{\sqrt{9+4E}}{4}
                           ],
                           [
                                \frac{3}{2}
                           ],\cos^2\beta\,
                        )\nonumber\\
         &+&        \frac{1/2}{
                                  \Gamma(\frac{5}{4}-\frac{\sqrt{4E+9}}{4})
                                  \Gamma(\frac{5}{4}+\frac{\sqrt{4E+9}}{4})
                              }  
                       F(\,[
                             -\frac{1}{4}+\frac{\sqrt{9+4E}}{4},
                             -\frac{1}{4}-\frac{\sqrt{9+4E}}{4}
                           ],
                           [
                             \frac{1}{2}
                           ],\cos^2\beta\,
                        )\label{Eigenvalues}
\end{eqnarray}
}\\[-1ex]
This equation can be analyzed without too much effort. The zeros 
of the graph on the right hand side in eq.~(\ref{Eigenvalues}) are 
the odd eigenvalues of $H_1$ or the even eigenvalues of $H_2$, respectively. 
The graph for $\beta=1/2$ is shown in Fig.~\ref{figureWF3halfH1}.\\
\parbox{\textwidth}{\vspace{0.2cm}
  \refstepcounter{figure}
  \label{figureWF3halfH1}
  \begin{center}
  \begin{turn}{-90}
  \makebox[5cm]{
     \epsfxsize=5cm
     \epsfysize=5cm
     \epsfbox{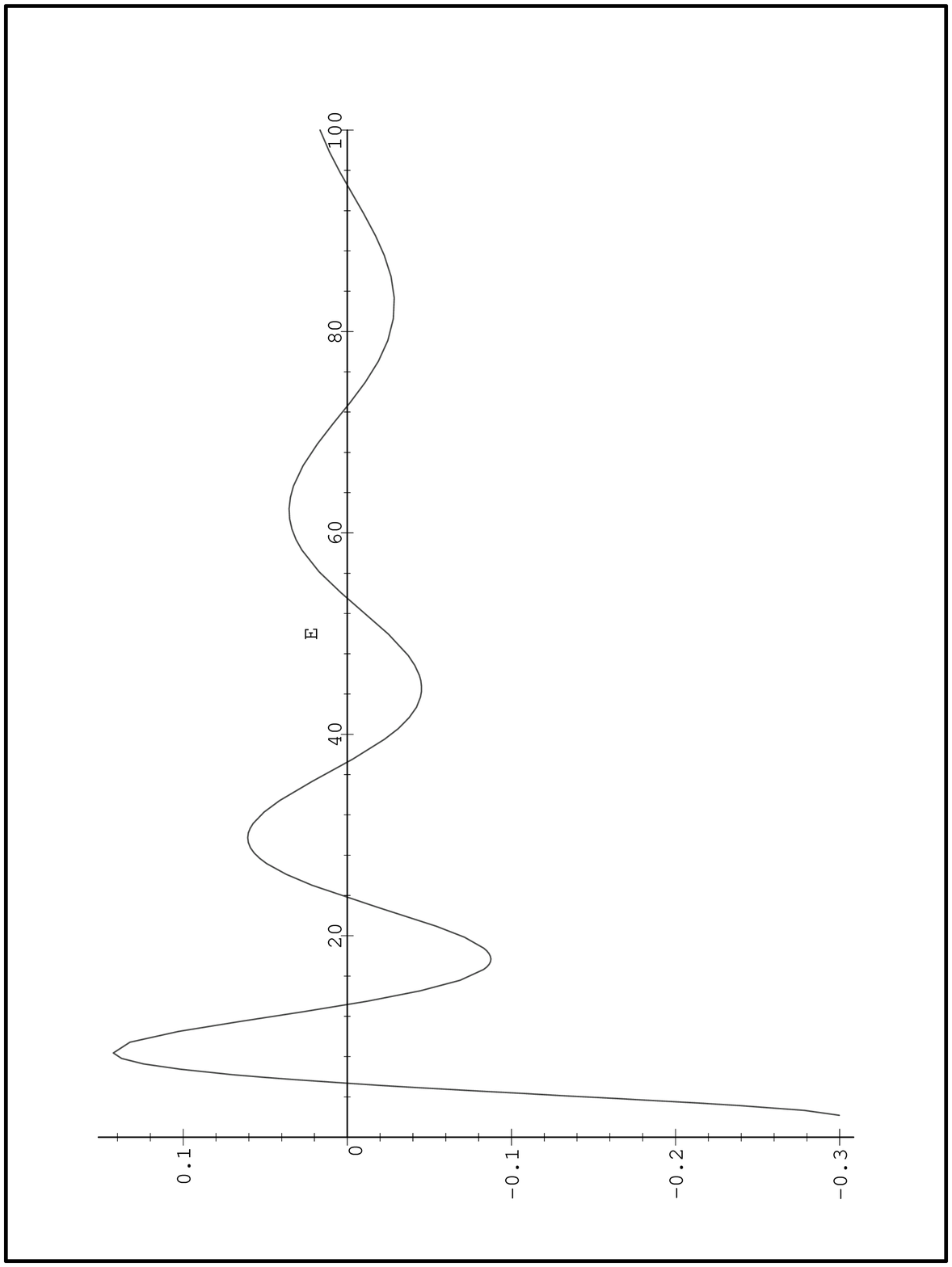}
  }
  \end{turn}
  \end{center}
  \center{{\bf Fig.{\thefigure}.} Odd eigenvalues of $H_1$ for 
                                  $\beta=\frac{1}{2}$}
}\\[1ex]
In Tab.~\ref{TabelleEigenwerte} we have collected some of the odd 
eigenvalues of $H_1$ for different values of $\beta$. If $\beta=\pi/2$ 
we obtain half of the integer spectrum of an algebraically solvable 
self similar potential. For very small $\beta$ we see the 
`almost zero mode'. 
\begin{center}
\refstepcounter{table}
\label{TabelleEigenwerte}
\begin{tabular}{|c|c|c|c|}
\hline
  & & & \\[-1ex]
n & $E_n^{\beta\,=\,0.1}$&$E_n^{\beta\,=\,0.5}$&$E_n^{\beta\,=\,\pi/2}$\\[1ex]
\hline
  &                      &               &     \\[-1ex] 
1 &   0.01471523         &  0.32986726   & 4   \\
3 &   4.07116545         &  5.35248438   & 18  \\
5 &  10.19138101         & 13.16289911   & 40  \\
7 &  18.39233651         & 23.78670947   & 70  \\
9 &  28.68676887         & 37.23243928   & 108 \\[1ex]
\hline
\end{tabular}
\center{{\bf Tab.~\thetable{}:}~ Some eigenvalues}
\end{center}
It is quite interesting to emphasize that for very small $\beta$ the 
eigenvalues converge to the spectrum of a self similar potential, too. 
The reason for this is the $\beta$-dependence of the shape of the potentials.
In Fig.~\ref{figurePotBeta05} and Fig.~\ref{figurePotBeta01} we have shown 
the basic structure of the potential $V_1$ for $\beta=0.5$ and $\beta=0.1$.
As $\beta$ approaches zero the potential splits into two independent 
self similar potentials with integer spectrum $E_n\,=\,n\,(n+3)$.\\  
\parbox{\textwidth}{\vspace{0.2cm}
\parbox{0.5\textwidth}
{
  \refstepcounter{figure}
  \label{figurePotBeta05}
  \begin{center}
  \begin{turn}{-90}
  \makebox[5cm]{
     \epsfxsize=5cm
     \epsfysize=5cm
     \epsfbox{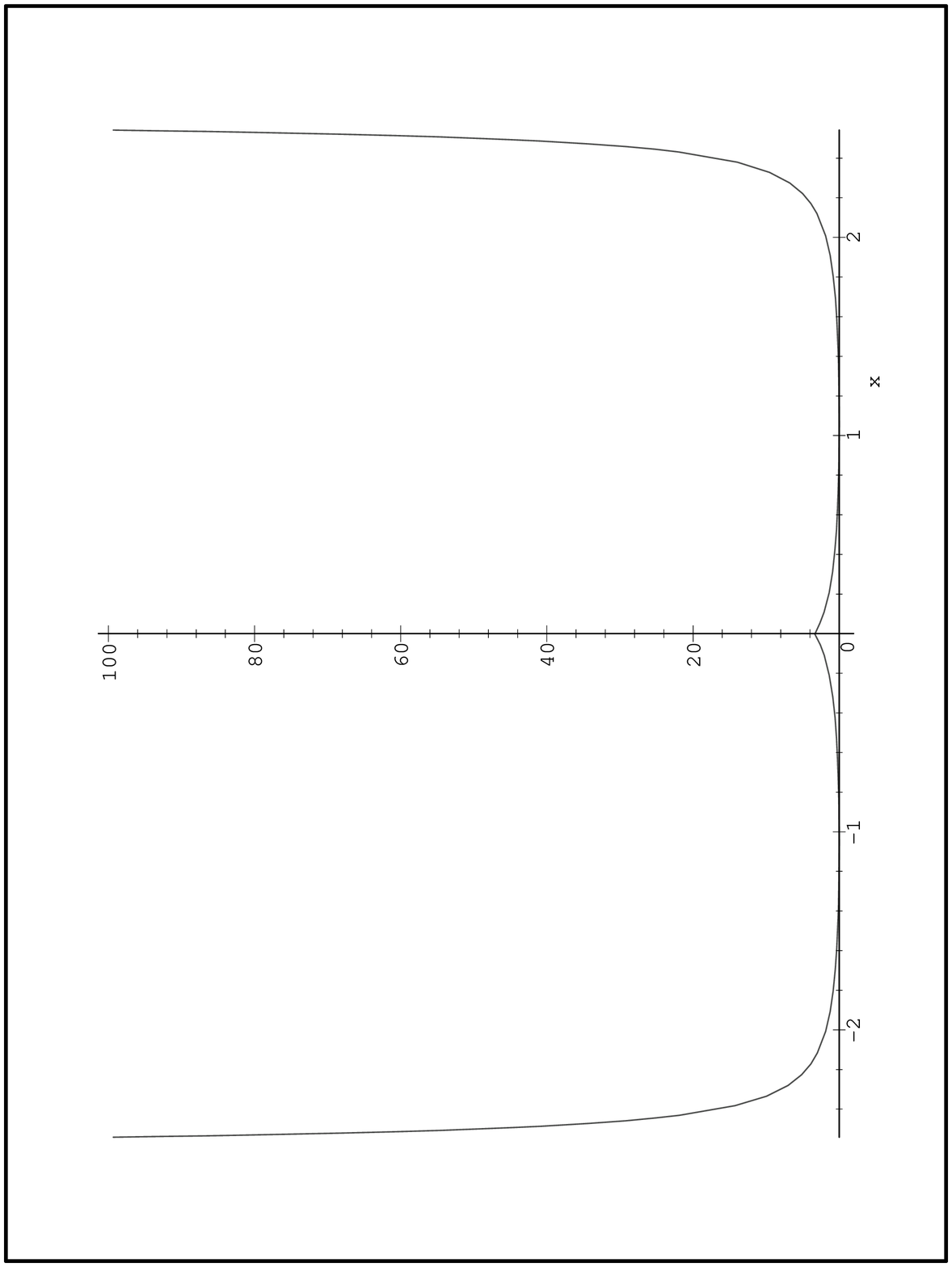}
  }
  \end{turn}
  \end{center}
  \center{{\bf Fig.{\thefigure}.} $V_1$ for $\beta=0.5$}
}\hfill
\parbox{0.5\textwidth}
{
  \refstepcounter{figure}
  \label{figurePotBeta01}
  \begin{center}
  \begin{turn}{-90}
  \makebox[5cm]{
     \epsfxsize=5cm
     \epsfysize=5cm
     \epsfbox{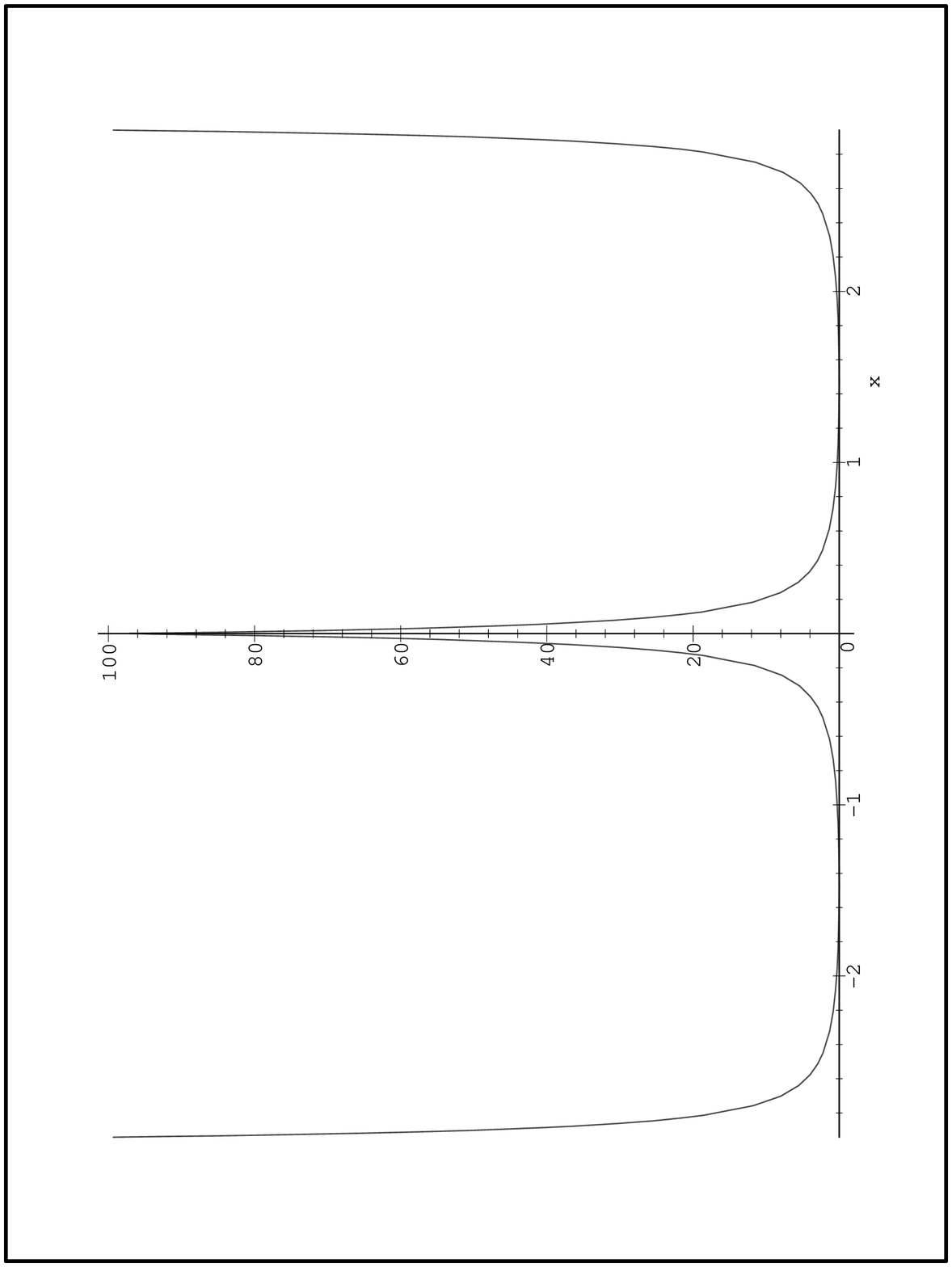}
  }
  \end{turn}
  \end{center}
  \center{{\bf Fig.{\thefigure}.} $V_1$ for $\beta=0.1$}
}
}

\subsection{Obtaining solutions of $H_2$}

The back transformation follows the pattern sketched in 
Fig.~\ref{figureSpektren}. The eigenvalues and eigenfunctions 
of the Hamiltonian $H_2$ can be obtained from those of $H_1$ by:
\begin{eqnarray*}
    \psi_n^{(2)} ~=~ \frac{1}{\sqrt{E_{n+1}^{(1)}}}\;A\psi_{n+1}^{(1)}
    \hspace{10ex}
    E_{n}^{(2)}  ~=~ E_{n+1}^{(1)} 
\end{eqnarray*}
In the next two figures we show the lowest normalized eigenfunctions.  
In Fig.~\ref{figureWF3halfOdd} we pictured only the first odd eigenfunction
of $H_1$, since for all higher ones there is no characteristic change in 
shape. In Fig.~\ref{figureWF3halfEven} we show the two lowest even 
eigenfunctions of the  Hamiltonian $H_2$.\\[-0.5ex] 
\parbox{\textwidth}{\vspace{0.2cm}
\parbox{0.5\textwidth}
{
  \refstepcounter{figure}
  \label{figureWF3halfOdd}
  \begin{center}
  \begin{turn}{-90}
  \makebox[5cm]{
     \epsfxsize=5cm
     \epsfysize=5cm
     \epsfbox{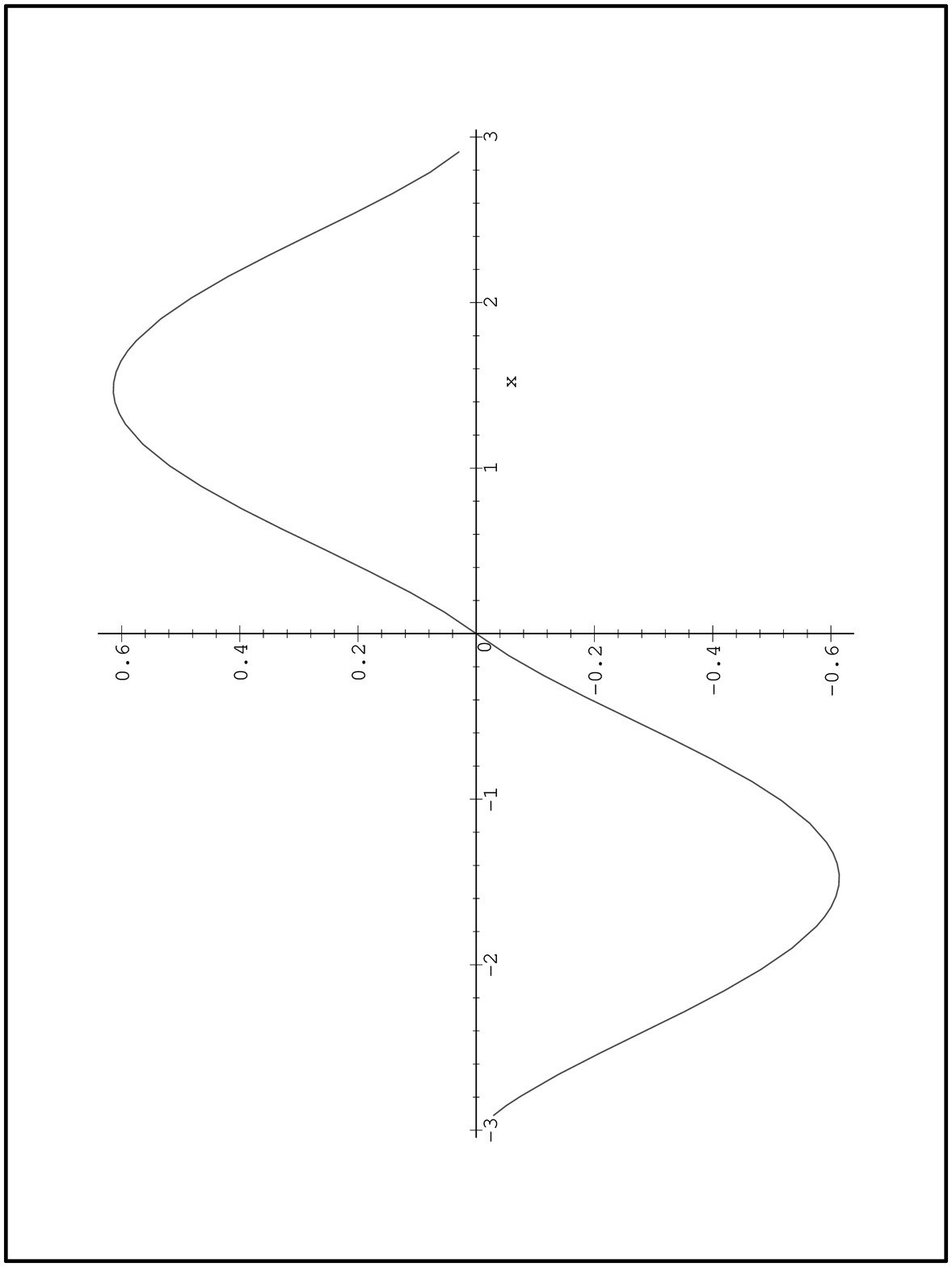}
  }
  \end{turn}
  \end{center}
  \center{{\bf Fig.{\thefigure}.} Odd eigenfunctions of $H_1$}
}\hfill
\parbox{0.5\textwidth}
{
  \refstepcounter{figure}
  \label{figureWF3halfEven}
  \begin{center}
  \begin{turn}{-90}
  \makebox[5cm]{
     \epsfxsize=5cm
     \epsfysize=5cm
     \epsfbox{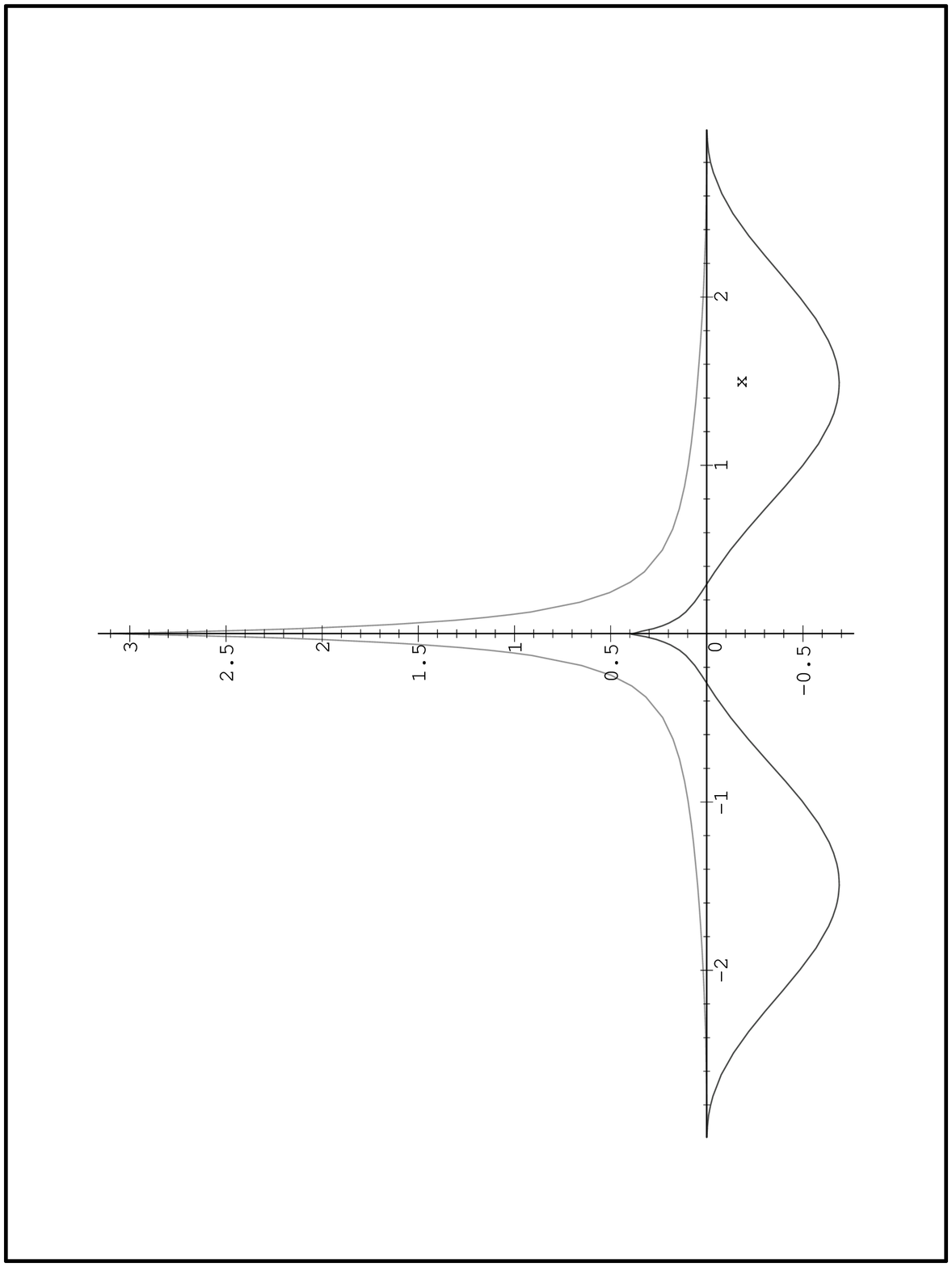}
  }
  \end{turn}
  \end{center}
  \center{{\bf Fig.{\thefigure}.} Even eigenfunctions of $H_2$}
}
}

\subsection{The lowest Eigenvalue}

To estimate the lowest eigenvalue $E_0$ in eq.~(\ref{Eigenvalues}) more 
precisely, we try to find an expansion of $E_0$ in terms of $\beta$. 
The expansion is around the small quantities $E$ and $~1\,-\,\cos^2\beta~$.
We start by expanding the square roots in eq.~(\ref{Eigenvalues}) 
and obtain:
\begin{eqnarray*}
   0 &=&  2\,\cos\beta\,F(\,[\,-1/2-E/6,\,1 +E/6\,],\,
                            [\,3/2\,],\,\cos^2\beta)\,
          \Gamma(2+E/6)\,\Gamma(1/2-E/6)\nonumber\\
      && + F(\,[\,1/2+E/6,\,-1-E/6\,],\,
               [\,1/2\,],\,\cos^2\beta\,)\,
               \Gamma(-E/6)\,\Gamma(3/2+E/6)
\end{eqnarray*}
Now we careful rewrite each single term as an expansion in the small 
quantities. The first term we consider is the hypergeometric function,
which appears in the second term of eq.~(\ref{Eigenvalues}). We are looking 
for a power expansion in the deviation $E$ of the parameters. This leads to:
\begin{eqnarray*}
  F(\,[\,1/2+E/6,\,-1-E/6\,],\,[\,1/2\,],\,x\,)~=~(1-x)
      ~+~\left[\,-\frac{x}{3}+\frac{(1-x)}{6}\ln|x-1|\,\right]\cdot E
      ~+~{\mathcal{O}}(E^2)
\end{eqnarray*}
Inserting this into the equation above we obtain:
\begin{eqnarray*}
  0 &=& 2\,\cos\beta\,F([-1/2-E/6,1+E/6],[3/2],\cos\beta^2\,)\,
        \Gamma(2+E/6)\,\Gamma(1/2-E/6)\\
    &+&
        \left[\,
               \sin^2\beta ~+~ E\,\left(\,
                                        -\frac{\cos^2\beta}{3}\,+\,
                                         \frac{\sin^2\beta}{6}\,
                                         \ln\sin^2\beta\,
                                  \right)\,
        \right]\,\Gamma(-E/6)\,\Gamma(3/2+E/6)
\end{eqnarray*}
To deal with the $\Gamma$-functions, we use the two expansions: 
\begin{eqnarray*}
   \Gamma(E/6)\Gamma(1/2-E/6) &=&\phantom{-}\,\frac{6}{E}\,\sqrt{\pi} 
                              ~+~ 2\,\ln 2\,\sqrt{\pi}
                              ~+~{\mathcal{O}}(E)\\
   \Gamma(-E/6)\Gamma(1/2+E/6)&=&-\,\frac{6}{E}\,\sqrt{\pi} 
                              ~+~ 2\,\ln 2\,\sqrt{\pi}
                              ~+~{\mathcal{O}}(E)
\end{eqnarray*}
Since we are interested in the limit $E\rightarrow 0$, the dominant 
contribution comes from the first terms. Using this to further simplify 
the transcendental equation  we obtain:
\begin{eqnarray*}
  0 &=& 2\,\cos\beta\,F([-1/2-E/6,1+E/6],[3/2],\cos\beta^2\,)\,
        (1+E/6)\\
    &-&
        \left[\,
               \sin^2\beta ~+~ E\,\left(\,
                                        -\frac{\cos^2\beta}{3}\,+\,
                                         \frac{\sin^2\beta}{6}\,
                                         \ln\sin^2\beta\,
                                  \right)\,
        \right]\,(3+E)/E
\end{eqnarray*}
The last term to consider is the remaining hypergeometric function. 
In principle we should treat this case completely analogous to the 
case before. But there is a problem in doing that. It is not obvious 
what the closed expression for the series in $x$ to the order 
${\mathcal{O}}(E)$ would be. All what can be done is to write down 
this expansion. By nice circumstances it is possible to find the value 
of this series at $x=1$, i.e. the sum of all its coefficients:
\begin{eqnarray*}
  F([-1/2-E/6,1+E/6],[3/2],\,x\,) 
  &=& \underbrace{F([-1/2,1],[3/2],\,x\,)}_{\hbox{elementary fct.}}\\ 
  &-& \left(\vbox{\vspace{2ex}}\right.
            \underbrace{
                         \frac{1}{6}\,x+\frac{1}{60}\,x^2
                         +\frac{1}{180}\,x^3+\frac{61}{22680}\,x^4\ldots
                       }_{
                         \hbox{$\frac{1}{12}\,+\,\frac{1}{6}\,\ln\,2~$
                               if $~x\,=\,1$}
                       } 
      \left.\vbox{\vspace{2ex}}\right)\cdot E ~+~ {\mathcal{O}}(E^2)
\end{eqnarray*}
But now our limit helps again. Since the term, not known precisely, 
contributes to the third power in $E$, we can neglect its effect in 
the expansion to lowest order in $E$. 
In the equation obtained after this 
manipulations  $E$ and $\beta$ decouple completely and we are able to 
solve  for $E_0$. The resulting expression can be expanded to extract 
the right power of $\beta$ to lowest order. This leads to:  
\begin{eqnarray}\label{Asympt}
   E_0\,(\beta) &=&  \frac{3}{2}\,\beta ^{2} ~+~ {\mathcal{O}}(\beta^{4})
\end{eqnarray}
Some short remarks. The question arises: is it sensible to take the 
limit $\beta\rightarrow 0$ while keeping $\alpha$ fixed at the same time, 
i.e. is $E_0\,=\,0$ a solution in this limit ? From the formula above 
it seems to be true. But in fact this limit is not a continuous one. The 
resolution of this puzzle is contained in what we had discussed before. 
If one looks at the potentials in Fig.~\ref{figurePotBeta05} and 
Fig.~\ref{figurePotBeta01}, than all mystery disappears. 
In the limit $\beta\rightarrow 0$ the barrier of the potential at $x=0$ 
grows  to infinity. Thus a wave function localized at $x=0$ has to 
vanish discontinuously. Is $\beta$ small but finite the corresponding 
wave function, the `almost zero mode', exists.\\ 

\noindent
Last but not least we want to discuss the precision of our 
approximation, which can be determined by comparing the exact 
with the estimated lowest eigenvalues $E_0(\beta)$. This is done in 
Fig.~\ref{figureVergleich}.
The upper curve corresponds to the asymptotic formula. The deviation 
in the interesting limit\linebreak 
($E$ and $\beta$ small) is around 1 percent, which is quite good.\\
\parbox{\textwidth}{\vspace{0.2cm}
  \refstepcounter{figure}
  \label{figureVergleich}
  \begin{center}
  \begin{turn}{-90}
  \makebox[5cm]{
     \epsfxsize=5cm
     \epsfysize=5cm
     \epsfbox{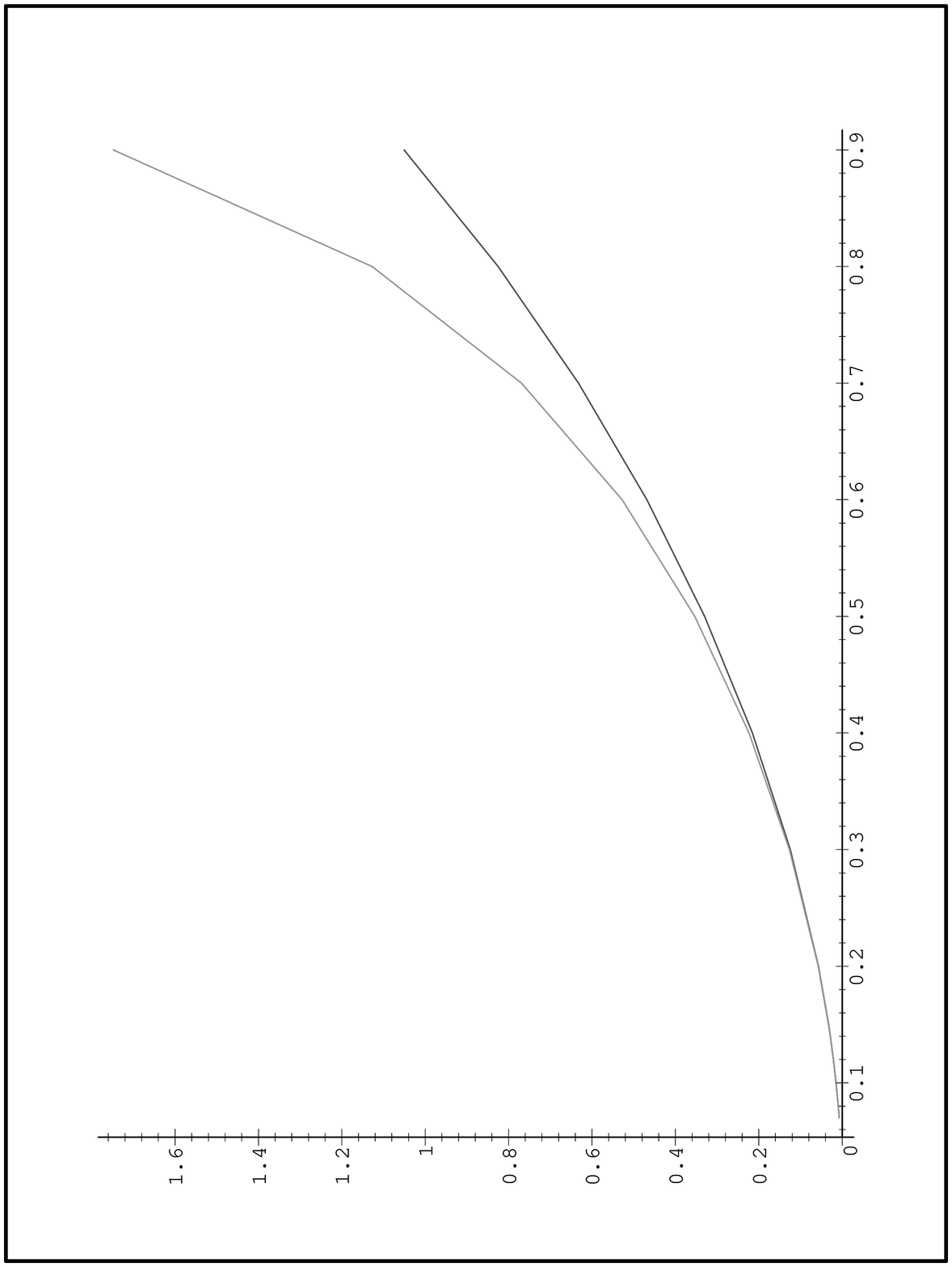}
  }
  \end{turn}
  \end{center}
  \center{{\bf Fig.{\thefigure}.} The precision of the asympt. formula}
}\\[1ex]
We conclude with the substitution of the technical $\alpha$ and 
$\beta$ parameters defined in section \ref{SectionIntro} by the 
physical parameters and undo the scaling of $E$ as introduced at 
the beginning of subsection \ref{SubSectionDGL}. Then we 
obtain the power law for the lowest eigenvalue in the form:
\begin{eqnarray*}
   m_0^2 ~=~ \frac{3}{2}\;|\Lambda|^2\,x_0^2
         ~+~ {\mathcal{O}}(|\Lambda|^{3})
\end{eqnarray*}  
From the pure scaling argument of subsection \ref{SubSectionDGL} 
which applies to the spectrum of the selfsimilar cousin of the 
class of  Hamiltonians considered here one would expect a linear
dependence of the cosmological constant. But now we find that 
the true dependence is a quadratic one.

\newpage


\vskip0.5cm
\noindent{\bf Acknowledgements:}

\noindent 
We want to thank Andreas Karch and Lisa Randall for suggesting this 
problem and useful discussions during the whole computation. 
Furthermore I would like to thank Axel Krause and Antonios 
Papazoglou for valuable comments.


\bibliographystyle{utphys}
\bibliography{engl}

\end{document}